\documentclass[epsfig,12pt]{article}

\usepackage[dvips]{graphicx}
\usepackage{graphicx}
\usepackage{epsfig}

\title{\hfill {\small ULB-TH-04/28} \\
\hfill {\small ITEP-PH-5/2004} \\
~~~ \\
Mass and decays of  Brout-Englert-Higgs scalar with extra generations}
\author{J.-M. Fr{\`e}re\\
Physique Th\'eorique, ULB, Brussels\\
  A.N. Rozanov \\
ITEP, Moscow, Russia\\
and\\
CPPM-IN2P3-CNRS-Universite Mediterrane, Marseille, France\\
M.I. Vysotsky \\
ITEP, Moscow, Russia}

\begin{document}
\maketitle

\begin{abstract}

The higher bound on the mass of the Brout-Englert-Higgs scalar boson
(BEH boson for brevity) arising from radiative corrections is not stable when
the Standard Model is extended to include non-decoupling particles.
In particular additional generation{s} of fermions allow for a heavier scalar.
We investigate how the decay branchings of scalar boson are affected by the opening
of new channels.
\end{abstract}

The precision measurement of $Z$-boson parameters, $W$-boson mass,
top quark mass and the  value of the running electromagnetic coupling
constant at the $Z$-boson mass allow to predict the mass
of the Brout-Englert-Higgs (BEH)
boson in the framework of the Standard Model.

\begin{table}[t]
\begin{center}
\caption{LEPTOP fit of the precision observables.}\label{tab:observables}

\begin{tabular}{|l|l|l|r|}
\hline
Observ. & Exper.  & LEPTOP & Pull \\
           &  data   & fit    &      \\
\hline
$\Gamma_Z$ {\tiny[GeV]} &    2.4952(23) &  2.4969(15)  & -0.7   \\
$\sigma_h$ [nb] &   41.540(37)& 41.481(14)  & 1.6    \\
$R_l$ &   20.767(25)& 20.736(18)  & 1.2   \\
$A_{FB}^l$ & 0.0171(10)  &  0.0164(2)  & 0.7   \\
$A_{\tau}$ & 0.1465(33) &  0.1480(11)  & -0.5   \\
$R_b$ &    0.2164(7) &   0.2156(1)  & 1.2   \\
$R_c$    &    0.172(3)&   0.1723(1)  & -0.1   \\
$A_{FB}^b$  &    0.0997(16)&   0.1038(8)  & -2.6   \\
$A_{FB}^c$  &    0.0706(35)&   0.0742(6)  & -1.0   \\
$s_l^2$ {\tiny $(Q_{FB})$}   &    0.2324(12)  &   0.2314(1)  & 0.8   \\
\hline
$s_l^2$ {\tiny ($A_{LR}$)} & {\it 0.2310(3)}&  {\it  0.2314(1)}  & -1.6   \\
$A_b$ &  0.925(20) &  0.9348(1)  & -0.5   \\
$A_c$ &  0.670(26)&   0.6683(5)  & 0.1   \\
\hline
$m_W$ {\tiny [GeV]} & 80.426(34) & 80.391(20)  & 1.0   \\
\hline
$m_t$ {\tiny [GeV]}    & 178.0(4.3) &   { 177.5(3.9)} & 0.1\\
$M_{BEH}$ {\tiny [GeV]}    &   &  { $103^{+54}_{-39} $} &  \\
$\hat{\alpha}_s$ &           &  0.1183(27) &  \\
$\bar{\alpha}^{-1}$ & 128.936(49)           & 128.924(48) & 0.2  \\
{\small $\chi^2/n_{dof}$} & & 16.7/12 & \\
\hline
\end{tabular}
\end{center}
\end{table}

Table \ref{tab:observables} contains the
LEPTOP fit of electroweak observables to the experimental data
updated in spring 2004. The increase
 of the $t$-quark mass according
to the D0 reanalysis pushes the scalar boson mass up:
\begin{equation}
(M_{BEH})^{\rm 2004}_{\rm Standard ~ Model} = (103 + 54 - 39) \; {\rm
GeV} \;\; , \label{1}
\end{equation}
and the quality-of-fit is good:
\begin{equation}
\;\;\;\;\;\;\;\; \chi^2/n_{\rm d.o.f.} = 16.7/12 \;\; .
\label{2}
\end{equation}
So, in the framework of the 3-generations Standard Model, the central value of $M_{BEH}$ is
close to the direct-search lower bound from LEP II, $M_{BEH} > 115$
GeV. Most probably it should be lighter than 200 GeV. This
prediction  originates from the
electroweak radiative corrections, which in first
approximation look like:
\begin{equation}
\delta V_i \approx (m_t/M_Z)^2 -  s^2 \ln(\frac{M_{BEH}}{M_Z})^2
\;\; , \label{3}
\end{equation}
where $s \equiv \sin\theta$, $\theta$ is the electroweak mixing angle,
while the definitions of functions $V_i$ are given in \cite{1}. It is
evident that introducing heavy fermionic generations will add new
contributions to the first term on right hand side of eq.(\ref{3})
and in order for functions $V_i$ to remain the same $M_{BEH}$ should
become larger. Indeed, it was noted in paper \cite{2} that a 500
GeV scalar does not contradict electroweak precision data if
accompanied by a fourth generation. In paper \cite{3}, the
five-dimensional parameter space of the model with a fourth generation
($M_{BEH}, m_U, m_D, m_E, m_N$) was investigated and the values of new
quark and lepton masses which allow for a heavy scalar were obtained.

In the present paper we investigate how the BEH-scalar decay branching
ratios change when new generations are added to the Standard
Model. Let us first determine how many new generations can be added.
New quarks contributions to the radiative
corrections depend mainly on the difference of the masses of up
and down quarks. Concerning leptons the difference of masses of
neutral and charged isodoublet members matters as well, however  a new
aspect emerges: the $\chi^2$  diminishes rapidly
when the new neutral lepton mass approaches one half of the $Z$-boson mass,
$m_N \approx 50$ GeV.

\begin{figure}[t]
\centering
\epsfig{file=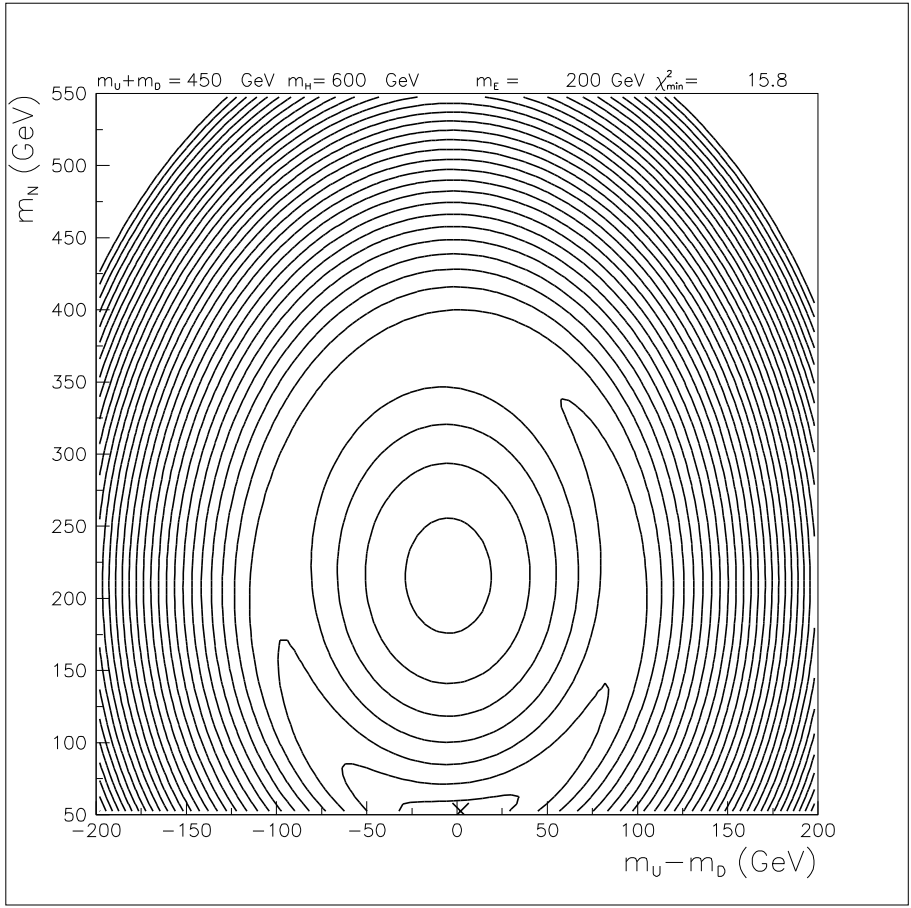,
width=12cm,height=12cm,angle=0}
\caption{ Exclusion plot for one extra generation. }
\label{WW2Fermi}
\end{figure}

To perform our fit we take initially a "typical value" $M_{BEH} = 600$ GeV,
$m_U + m_D = 450$ GeV and
$m_E = 200$ GeV and allow $m_U - m_D$ and $m_N$ to vary. If  $N_g$ is the
number of extra generations, for  $N_g =1$ , we obtain the
exclusion plot shown in Fig. 1. The minimum of $\chi^2$ is
at $m_N = 52.5$ GeV and almost degenerate $U$- and $D$-quarks,
$m_U - m_D = 2$ GeV, and it is equal to:
\begin{equation}
N_g = 1: \;\;\;\;\;\;\;\; \chi^2/n_{\rm d.o.f.} = 15.8/11 \;\; ,
\label{4}
\end{equation}
so the quality of fit is the same as that of the Standard Model, eq.
(\ref{2}).

\begin{figure}[t]
\centering
\epsfig{file=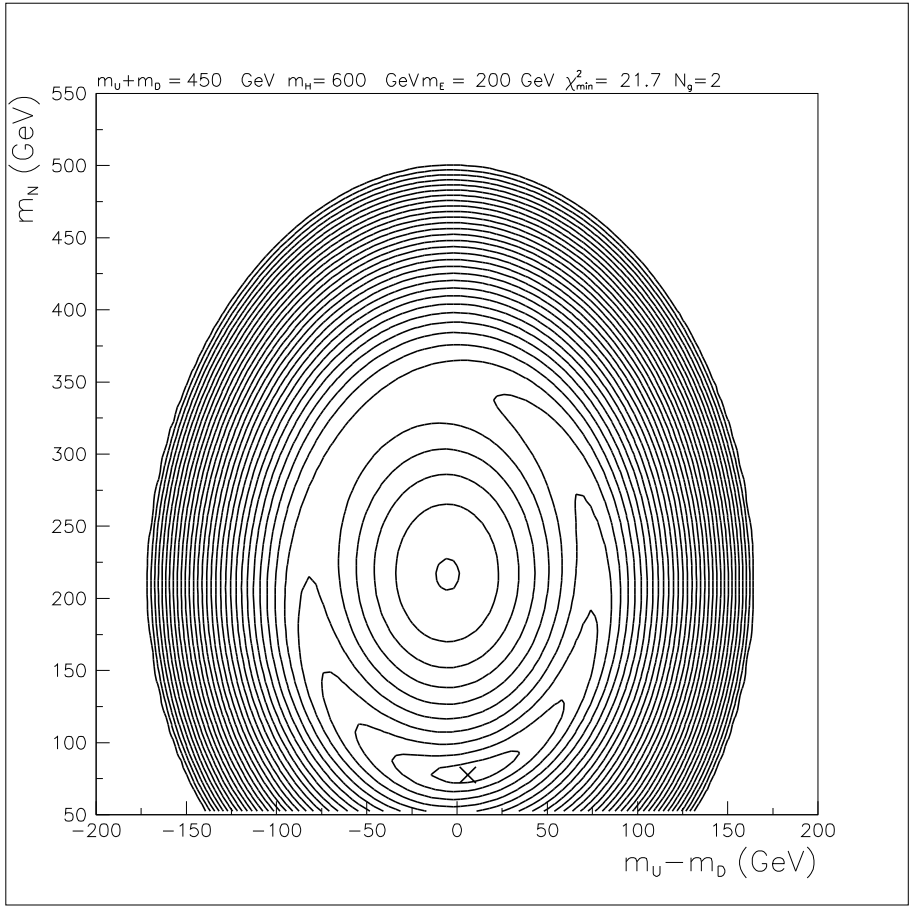,
width=12cm,height=12cm,angle=0}
\caption{ Exclusion plot for two extra generations.}
\label{WW2Fermi}
\end{figure}

In Fig. 2 the exclusion plot is given for $N_g =2$. In case of 2
extra generations we take masses of new fermions to be degenerate:
$m_{U_4} = m_{U_5} = m_U$, $m_{D_4} = m_{D_5}= m_D$, $m_{E_4} =
m_{E_5} = m_E$, $m_{N_4} = m_{N_5} = m_N$ in order to avoid
a multiplication of the number of fitted parameters. While we have
not exhausted the full parameter space, some other attempts did not
give more encouraging results. At $\chi^2$ minimum up and
down quarks are still degenerate, like in case of one extra
generation. Neutral leptons become a little bit heavier, $m_N
\approx 60$ GeV., but the quality of fit  worsens considerably:
\begin{equation}
N_{g=2}: \;\;\;\;\;\;\;\; \chi^2/n_{\rm d.o.f.} = 21.7/11
\label{5}
\end{equation}

We analyze the case with 3 extra generations in the same manner as
that of 2 extra generations, taking new quarks and leptons with
the same isospin projection degenerate and allow $m_N$ and $m_U -
m_D$ to vary. We get:
\begin{equation}
N_g = 3: \;\;\;\;\;\;\;\; \chi^2/n_{\rm d.o.f.} = 30.4/11 \;\; ,
\label{6}
\end{equation}
and comparing with eq.(\ref{4}) we conclude that 3 extra
generations with such parameters are excluded at the level of 4 standard deviations.

So, not more than two extra fermionic generations (at least with the mass pattern
considered) are allowed by current precision data.

The BEH scalar boson widths to  pairs of intermediate vector bosons are
given by the following formulas:
\begin{equation}
\Gamma_{BEH\to WW} = \frac{\alpha M_{BEH}^3}{16 s^2 M_W^2}\left[1-
(\frac{2M_W}{M_{BEH}})^2+12(\frac{M_W}{M_{BEH}})^4\right]\left[ 1-
\frac{4M_W^2}{M_{BEH}^2}\right]^{1/2}\;\; ,  \label{7}
\end{equation}
\begin{equation}
\Gamma_{BEH\to ZZ} = \frac{\alpha M_{BEH}^3}{32 s^2 c^2 M_Z^2}
\left[1-(\frac{2M_Z}{M_{BEH}})^2+12(\frac{M_Z}{M_{BEH}})^4\right]
\left[1-\frac{4M_Z^2}{M_{BEH}^2}\right]^{1/2}
\;\; ,  \label{8}
\end{equation}
and they rapidly increase when the scalar particle becomes heavier.

The decay widths to fermion pairs read:
\begin{equation}
\Gamma_{BEH\to f\bar f} = N_c
\frac{M_{BEH}}{8\pi}(\frac{m_f(M_{BEH})}{\eta})^2\left[1-\frac{4m_f(M_{BEH})^2}
{M_{BEH}^2}\right]^{3/2}\;\; ,  \label{9}
\end{equation}
where $N_c =1$ for lepton, $N_c =3$ for quarks and $\eta$ is the scalar
boson expectation value. In case of BEH decay to quark-antiquark pair
QCD running of quark mass takes into account leading gluonic corrections
to $BEH - q \bar{q}$ vertex.
An extra factor $1-\frac{4m_f^2}{M_{BEH}^2} = v_f^2$ ($v_f$ is the fermion
velocity) is due to the fact that Dirac fermions are produced in
$P$-wave when a scalar particle decays. In order to compare the
width of fermionic decays to the width of the vector bosons channel, it is
convenient to present eq.(\ref{9}) in the following way:
\begin{equation}
\Gamma_{BEH\to f\bar f} = N_c \frac{\alpha M_{BEH}}{8s^2}(\frac{m_f(M_{BEH})}{M_W})^2\left[1-\frac{4m_f(M_{BEH})^2}
{M_{BEH}^2}\right]^{3/2}
\;\; .  \label{10}
\end{equation}

As long as the BEH boson decays to the pair of vector bosons is
kinematically forbidden ($M_{BEH} < 160$ GeV), fermionic decays
dominate. In the case of the 3-generations Standard Model it is the decay to $b\bar b$
pair, see \cite{4}. The BEH branchings in the Standard Model, calculated using
HDECAY program \cite{105} are shown on Fig.3.
If extra generations are present and if the decay
to the pair of heavy neutral leptons is kinematically allowed, then it
dominates \cite{5}. It can be seen on the Fig.4 for the case $m_N= 53$ GeV,
$m_E=200$ GeV, $m_U=m_D=225$ GeV (calculated with modified version of
HDECAY).
The possibility of observing such an "invisible"
BEH boson decay at the LHC was discussed in a recent paper \cite{11}.

\begin{figure}[t]
\centering
\epsfig{file=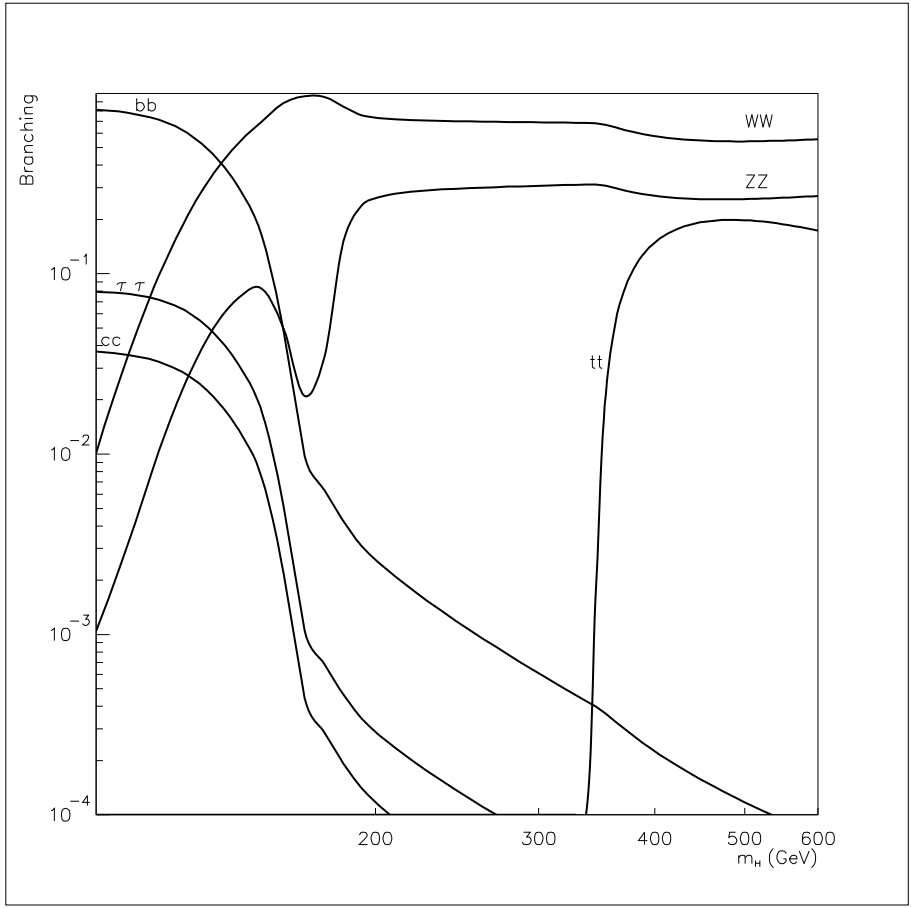,
width=12cm,height=12cm,angle=0}
\caption{ Branching ratios for scalar boson decays in Standard Model.}
\label{WW2Fermi}
\end{figure}

\begin{figure}[t]
\centering
\epsfig{file=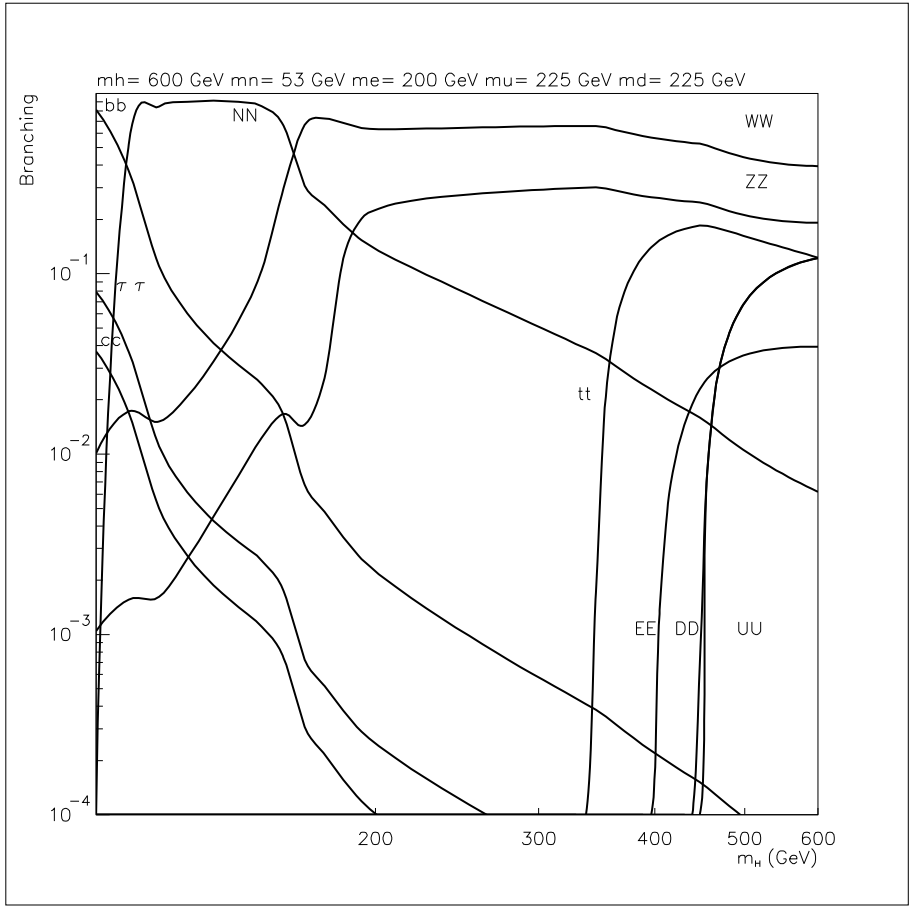,
width=12cm,height=12cm,angle=0}
\caption{ Branching ratios for scalar boson decays in case of one extra generation.}
\label{WW2Fermi}
\end{figure}

In the opposite case of very heavy BEH boson, $M_{BEH} = 600$ GeV,
the vector boson decay channel dominates in the strict Standard Model.
Let us
investigate if decays to new fermions can considerably diminish
this branching ratio. Let us adjust the mass of fermions
in such a way that the BEH boson width to this channel becomes maximal. According to
eq.(\ref{10}) this happens for $m_f^2 = M_{BEH}^2/10$:
\begin{equation}
(\Gamma_{BEH\to f\bar f})_{\rm max} \approx 0.09 N_c \frac{\alpha
M_{BEH}^3}{16 s^2 M_W^2} \;\; , \label{11}
\end{equation}
and for one extra degenerate fermion generation the sum over all
fermionic modes gives\footnote{It follows from the first part of
the paper that in order to allow a heavy BEH boson, masses in fermion
isodoublets should be split. We take however degenerate doublets here for
a quick estimation in
order to diminish the parameter space; it is clear that the pattern of
heavy BEH boson decays will not alter drastically if the splitting is
 taken into account.}:
\begin{equation}
\Gamma_{BEH\to {\rm new ~ degenerate ~ fermions}} = 0.09 \cdot 8
\frac{\alpha M_{BEH}^3}{16 s^2 M_W^2} \;\; . \label{12}
\end{equation}

Finally, the BEH boson decay to the pair $t\bar t$ should also be included :
\begin{equation}
\Gamma_{BEH\to t\bar t} = \frac{3\alpha M_{BEH}}{8
s^2}(\frac{m_t(M_{BEH})}{M_W})^2\left[1-\frac{4m_t(M_{BEH})^2}{M_{BEH}^2}\right]^{3/2}
\;\; .  \label{13}
\end{equation}
In order to calculate BEH boson width eq.(\ref{7} - \ref{9}) were used,
$\alpha/(s^2M_W^2)=\alpha/((sc)^2M_Z^2)=(\pi \eta^2)^{-1}$ were
substituted in  eq.(\ref{7} - \ref{8}), $m_f=
190$ GeV were taken in eq.( \ref{9}) and numerical value $
 \eta = 246$ GeV  was substituted.

For the case $M_{BEH} = 600$ GeV we obtain:

\begin{center}
\begin{tabular}{ccccc}
$\Gamma_{\rm new ~ degenerate ~ fermions}$ & $\Gamma_{WW}$ &
$\Gamma_{ZZ}$ & $\Gamma_{tt}$ & $\Gamma_{total}$ \\ 57 & 64 & 32 & 21 &
175
\end{tabular}
\end{center}

All widths are in GeV. In case of decays to quark pairs
QCD corrections which enhance width by approximately 20\%
are taken into account \cite{110}:

\begin{eqnarray}
\Gamma_{tt, QQ} & = & 3
\frac{M_{BEH}}{8\pi}(\frac{m_f(M_{BEH})}{\eta})^2[1+ 5.667
\alpha_s/\pi +(35.94 - 1.359 n_f)(\alpha_s/\pi)^2 + \nonumber \\ &
+ & (164.139 - 25.771 n_f + 0.259 n_f^2)(\alpha_s/\pi^3]
\left[1-\frac{4m_f(M_{BEH})^2} {M_{BEH}^2}\right]^{3/2}\;\; .
\label{14}
\end{eqnarray}

The doubling of the BEH boson width which occur in case of 2 extra
generations would lead to four times less events of BEH boson
decay to ``golden mode'' $BEH \to ZZ$ around the cross-section
maximum:
\begin{equation}
\sigma_{pp \to BEH \to ZZ} \sim \frac{\Gamma_{BEH \to ZZ} \Gamma_{BEH
\to gg}}{(E - M_{BEH})^2 + \Gamma_{BEH}^2/4} \;\; . \label{14}
\end{equation}

However since the $gg \to BEH$ fusion proceeds through quark triangles, new
quarks enhance it \cite{6}. In the Standard Model the triangle with top
quark dominates; as far as masses of new quarks are close to that
of top, the  production amplitude in case of two extra generations is
enhanced by factor 5 (contribution of $U_4, U_5, D_4, D_5$ should
be taken into account), hence , the production cross section is about 25 times that of
the Standard Model.

 We have thus seen that the  noncoherent effect of
heavy BEH boson($M_{BEH} \sim 500 - 600$ GeV) decay to new quarks
diminishes slightly the  effect of coherent enhancement of
BEH bosons production in gluon
fusion.
However, relatively light scalars ($M_{BEH} < 200$ GeV)
would be much more severely affected
by the now-predominant decay channel into  a pair of neutral leptons,
possibly  making such a BEH
invisible.
Consequences of the additional fermion families
on the BEH boson production at Tevatron and LHC was discussed in some details in
\cite{12}.

Acknowledgements

JMF wants to acknowledge the support of the IISN and of the Belgian
science policy office (IAP V/27);  MV acknowledge the support of the
program FS NTP FYaF 40.052.1.1.1112 and grant N SH - 2328.2003.2.

\end{document}